\documentclass[aps, twocolumn, prb]{revtex4-1}%
\usepackage{mathptmx} 
\usepackage{amsfonts}
\usepackage{amsmath}
\usepackage{amssymb}
\usepackage{graphicx}
\usepackage{bm}
\usepackage{natbib}
\usepackage{color}
\usepackage[colorlinks=true,linkcolor=red,anchorcolor=blue,citecolor=red, urlcolor=blue]{hyperref}

\begin{document}
\title{Determination of $g$-factor in InAs two-dimensional electron system by capacitance spectroscopy}\thanks{Published in Applied Physics Express on May 20, 2019. H. Irie, T. Akiho, and K. Muraki,
\href{https://doi.org/10.7567/1882-0786/ab1c7c}{Appl. Phys. Express {\bf12} 063004 (2019)}. This Accepted Manuscript is available for reuse under a CC BY-NC-ND 3.0 licence after the 12 month embargo period provided that all the terms and conditions of the licence are adhered to.}
\author{Hiroshi Irie, Takafumi Akiho, and Koji Muraki}
\affiliation{NTT Basic Research Laboratories, NTT Corporation, 3-1 Morinosato-Wakamiya, Atsugi 243-0198, Japan}
\date{\today}

\begin{abstract}
We determine the effective $g$-factor ($|g^\ast|$) of a two-dimensional electron gas (2DEG) using a new method based on capacitance spectroscopy.
The capacitance-voltage profile of a 2DEG in an InAs/AlGaSb quantum well measured in an in-plane magnetic field shows a double-step feature that indicates the Zeeman splitting of the subband edge.
The method allows for simultaneous and independent determination of $|g^\ast|$ and effective mass $m^\ast$.
Data suggest that the biaxial tensile strain in the InAs layer has considerable impacts on both $m^\ast$ and $g^\ast$.
Our method provides a means to determine $|g^\ast|$ that is complementary to the commonly used coincidence technique.

\end{abstract}
\maketitle

The electron $g$-factor is a fundamental quantity that governs the coupling between the spin degree of freedom and external magnetic fields.
In the solid state, the $g$-factor is altered from its value in vacuum ($g_e \approx 2$) by spin-orbit coupling and can even change sign, thus becoming a material-dependent parameter referred to as the effective $g$-factor, denoted as $g^\ast$.
This coupling, making $g^\ast$ dependent on various parameters such as electric field and quantum confinement, allows for controlling the spin degree of freedom by electrical means, which forms the basis for spintronics and quantum information processing.~\cite{Wu2010}
Recently, the combination of the strong spin-orbit interaction inherent in narrow-gap semiconductors and external magnetic fields has proven to provide routes to the emergent quantum phase,~\cite{Alicea2012,Suominen2017} where $g^\ast$ plays an important role.
In a two-dimensional electron gas (2DEG),  a common platform for various nanostructures and hybrid devices, $g^\ast$ is determined by a method known as the coincidence technique.~\cite{Fang1968,SmithIII1987,Nicholas1988,Brosig2000}
The method uses magnetic field $B$, with the angle $\theta$ from the sample normal varied to tune the ratio $r$ ($=|g^\ast|m^\ast/\hbar e \cos \theta$) between the Zeeman splitting $E_Z = |g^\ast| \mu_B B$ and cyclotron energy $\hbar e B_\perp/m^\ast$, where  $\mu_B$ is the Bohr magneton, $\hbar$ is Planck's constant divided by $2\pi$, $e$ is the elementary charge, $B_\perp = B \cos \theta$ is the perpendicular component of $B$, and $m^\ast$ is the effective mass.
Then $|g^\ast|$ is known from $\theta$ at which particular resistance minima disappear due to level coindicence.
However, in the presence of $B_\perp$, electron-electron interaction can affect $g^\ast$ in a manner dependent on the Landau-level filling factor.~\cite{Nicholas1988,Ando1974}
Therefore, alternative means to determine $|g^\ast|$ without $B_\perp$, which can provide complementary information, will be helpful.
In this study, we present a new method based on capacitance spectroscopy that does not require $B_\perp$ and determine $|g^\ast|$ of a 2DEG in a quantum well (QW) of InAs, a narrow-gap semiconductor with strong spin-orbit interaction and large $|g^\ast|$.

Our method is based on the measurement of quantum capacitance $c_Q = e^2  (d n_s / d \mu)$, which is proportional to the thermodynamic density of states (DOS) $\mathcal{D} = d n_s / d \mu$ of the 2DEG~\cite{Luryi1988,Ali2011}  [Fig.~1(a)].
Here, $\mu$ is the chemical potential, and $n_s$ is the electron density.
We apply a strong in-plane magnetic field $B_\parallel$ and measure the Zeeman splitting of the subband edge [Fig.~1(b)] that appears as a double step in the capacitance-voltage profile.
Figure~2(a) shows the equivalent-circuit representation of the system.
The differential capacitance $c_\text{FG}$ (per unit area) between the 2DEG and the front gate, separated by an insulator with dielectric constant $\epsilon_b$ and thickness $d_b$, can be described by a series sum of $c_Q$ and geometrical capacitance $c_G = \epsilon_b/d_b$ as
\begin{equation}
c_\text{FG}^{-1} = c_G^{-1} + c_Q^{-1}.
\label{c_FG}
\end{equation}
Using this equation along with $e (d n_s / d V_\text{FG}) = c_\text{FG}$, where $V_\text{FG}$ is the front-gate voltage, one can show that the relation
\begin{equation}
\mu = e \int_{V_\text{th}}^{V_\text{FG}} \left[ 1 - \frac{c_\text{FG}(V_\text{FG}^\prime)}{c_G} \right] dV_\text{FG}^\prime
\label{mu}
\end{equation}
holds between $\mu$ and $V_\text{FG}$,~\cite{Khrapai2007} where $V_\text{th}$ is the threshold voltage at which the 2DEG starts to accumulate.
This relation can be used to translate voltage into energy from the measured $c_\text{FG}$ vs $V_\text{FG}$.
This allows us to obtain $E_Z$ (and hence $|g^\ast|$) from the double-step feature in the capacitance-voltage profile.

\begin{figure}[ptb]
\includegraphics[width = 0.45\textwidth]{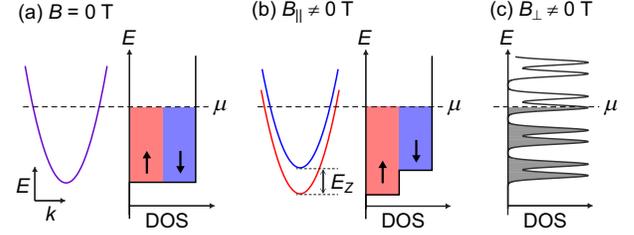}\caption{
Schematic illustrations of band dispersion and density of states of a two-dimensional electron gas subjected to (a) zero, (b) in-plane, and (c) perpendicular magnetic fields.}
\label{Fig1}
\end{figure}

\begin{figure}[ptb]
\includegraphics[width = 0.46\textwidth]{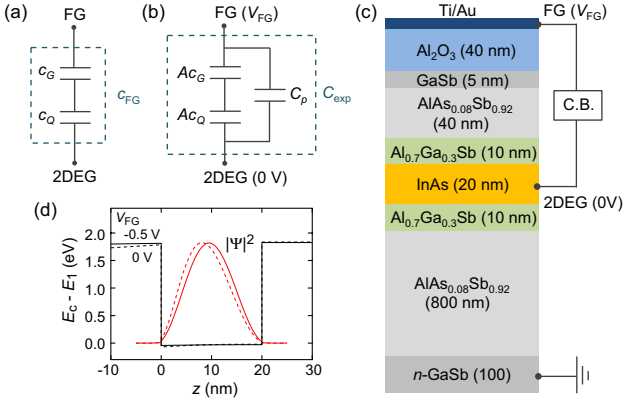}\caption{
(a) Equivalent-circuit representation of the differential capacitance $c_\text{FG}$ (per unit area) between the front gate and 2DEG.
(b) Equivalent circuit for differential capacitance $C_\text{exp}$ measured in the experiment.
(c) Layer structure of the device with a connection diagram for the capacitance-bridge (C.B.) measurement.
\textcolor{black}{
(d) Profiles of the conduction-band edge $E_c(z)$ measured from the lowest subband energy $E_1$ and squared envelope function $|\Psi(z)|^{2}$ calculated for $V_\text{FG} = -0.5$ (solid lines) and $0$~V (dashed lines).
}}
\label{Fig2}
\end{figure}

Measurements were conducted at $1.8$~K on a \textcolor{black}{square ($600~\mu\text{m} \times 590~\mu \text{m}$)} device fabricated from a heterostructure grown by molecular beam epitaxy on an $n$-type GaSb (001) substrate.
Figure~2(c) depicts the layer structure of the device.
The 2DEG is hosted in a 20-nm-wide InAs QW sandwiched by 10-nm-thick Al$_{0.7}$Ga$_{0.3}$Sb barriers.
The QW structure is flanked on both sides by outer AlAs$_{0.08}$Sb$_{0.92}$ barrier layers and capped with $5$-nm GaSb.
The AlAs$_{0.08}$Sb$_{0.92}$ layers were designed to lattice-match the GaSb substrate and thereby eliminate dislocation formation.
Note that the lattice constant of InAs (Al$_{0.7}$Ga$_{0.3}$Sb), $6.0501$ ($6.1139$)~\AA, is 0.52\% smaller (0.53\% larger) than that of GaSb ($6.0817$~\AA).
This induces a biaxial 0.52\% tensile (0.53\% compressive) strain in the InAs (Al$_{0.7}$Ga$_{0.3}$Sb) layer(s).
The device has two Ohmic contacts and a front gate with a $40$-nm-thick Al$_2$O$_3$ insulator atomic-layer deposited on the heterostructure.
We measured the capacitance $C_\text{exp}$ between the front gate and the 2DEG~\footnote{
Throughout this paper, we use the upper case $C$ to denote capacitance to distinguish it from capacitance per area $c$ denoted in the lower case.
} using a capacitance bridge (Andeen-Hagerling 2700A) as a function of $V_\text{FG}$ [Fig.~2(c)], at a frequency of $320$ Hz and an excitation voltage of $3$~mVrms.
\textcolor{black}{
The dissipation factor measured simultaneously with $C_\text{exp}$ was negligibly small for the $V_\text{FG}$ range studied, which confirms the irrelevance of charge trapping in the barrier layers.
}
The underlying $n$-GaSb (buffer layer and substrate) was electrically isolated from the 2DEG by the thick AlAs$_{0.08}$Sb$_{0.92}$ layer, and therefore did not contribute to $C_\text{exp}$.
Separate transport measurements on a Hall-bar device fabricated from the same wafer showed that the 2DEG had sheet density of $n_s = 3.65 \times 10^{15}$~m$^{-2}$ and low-temperature mobility of $50$~m$^2$/Vs at $V_\text{FG}=0$~V.
\textcolor{black}{
Self-consistent envelope-function calculations reveal a slightly asymmetric charge distribution within the QW with negligible penetration into the AlGaSb barriers for the $V_\text{FG}$ range relevant to the $|g^{\ast}|$ determination [Fig.~\ref{Fig2}(d)].
}
A magnetic field of up to $14$~T was applied either parallel or perpendicular to the 2DEG.

In actual experiments, parasitic capacitance $C_p$ exists, which enters in parallel [Fig.~2(b)].
In addition, $c_G$ is not perfectly constant and varies with $V_\text{FG}$ as shown below.
Therefore, we first show how we determined $C_p$ and $c_G$ by presenting the data obtained at zero magnetic field ($B = 0$~T) and in a perpendicular field $B_\perp$.
Figure~3(a) shows the measured $C_\text{exp}$ as a function of $V_\text{FG}$.
At $B=0$~T, $C_\text{exp}$ is nearly constant at $V_\text{FG} > \textcolor{black}{-0.8}$~V, and decreases sharply when the 2DEG is depleted at $V_\text{FG} \leq \textcolor{black}{-0.8}$~V.
As eq.~(\ref{c_FG}) shows, this sharp decrease represents the contribution of $c_Q$ to $c_\text{FG}$.
In the depletion region ($V_\text{FG} \leq \textcolor{black}{-0.8}$~V), $C_\text{exp}$ takes a finite value, corresponding to the parasitic capacitance, which mainly stems from the overlap between the front gate and Ohmic electrodes.

\begin{figure}[ptb]
\includegraphics[width = 0.35\textwidth]{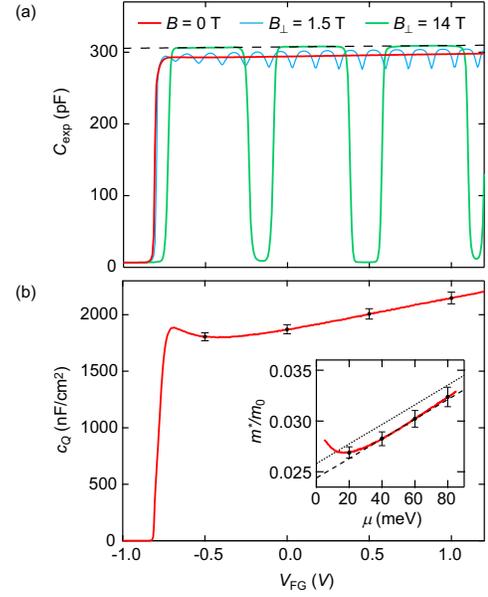}\caption{
(a) Capacitance-voltage profile measured at $B_\perp = 0$, $1.5$, and $14$~T.
The dashed line is a linear fit to the flattened top parts of the $B_\perp = 14$~T data, representing the geometrical capacitance $A c_G$.
(b) $c_Q$ obtained from $C_\text{exp}$ at $B = 0$~T.
Inset shows the effective mass calculated from $c_Q$ (solid line).
\textcolor{black}{
The error bars represent the maximum error caused by the linear approximation of $c_{G}$ vs $V_\text{FG}$.
}
The \textcolor{black}{dashed and dotted} lines are thoretical values for a $20$-nm-wide InAs QW with and without strain effects taken into account, respectively, taken from Ref.~\onlinecite{Lin-Chung1993}.
}
\label{Fig3}
\end{figure}

When a perpendicular field is applied, the DOS of the 2DEG splits into a series of peaks due to Landau quantization [Fig.~1(c)].
Accordingly, $C_\text{exp}$ oscillates as a function of $V_\text{FG}$, where the high (low) $C_\text{exp}$ indicates that the Fermi level lies within a Landau level (between Landau levels).~\cite{Smith1985}
Notably, maxima of $C_\text{exp}$ flatten out at a very high field ($B_\perp = 14$~T).
This happens because, at high fields, the large DOS of Landau levels makes  $c_Q$ much greater than $c_G$; consequently, the second term in eq.~(\ref{c_FG}) becomes negligible.~\cite{Yu2013}
Hence, we have $C_\text{exp} \approx A c_G + C_p$, where $A$ ($= \textcolor{black}{0.347}$~mm$^{2}$) is the area of the 2DEG.~\footnote{We estimated $A$ by comparing the density obtained by numerically integrating the measured capacitance and that determined from \textcolor{black}{the magneto-capacitance oscillations for $B_\perp = 1$--$2$~T at various $V_\text{FG}$}.
The \textcolor{black}{$2$}\% reduction from the lithographic size is likely due to overetching of the mesa side walls.}
In turn, this implies that, once $C_p$ is known, $c_G$ can be determined experimentally by measuring $C_\text{exp}$ at a sufficiently strong $B_\perp$ where $c_Q \gg c_G$.~\footnote{
A simple calculation shows that $c_Q/c_G \geq 100$ is satisfied for Gaussian Landau-level broadening of $\sigma \leq 2.4$~meV at $B_\perp = 14$~T for our sample geometry.
}
In the following analysis, we use the value $C_p = \textcolor{black}{6.8}$~pF, which we confirmed to be constant at high $B_\perp$.~\footnote{
This value agrees well with the minimum of $C_\text{exp}$ at $V_\text{FG} = 0.5$~V, which corresponds to the integer quantum Hall effect at Landau-level filling factor 2, where the DOS at the Fermi level is expected to become minimum.
}

As shown by the dashed line in Fig.~3(a), the $C_\text{exp}$ values in the flat-top regions can be fitted by a single straight line.
The fit line has a finite slope, which indicates that $c_G$ slighty increases with $V_\text{FG}$.
This is because the 2DEG has a finite width, and its centroid $\langle z \rangle$ within the QW varies with $V_\text{FG}$, reflecting the change in the confinement potential.~\cite{Hampton1995, Ali2011}
This effect can be incorporated by representing $c_G$ as
\begin{equation}
c_G^{-1} = \frac{d_b}{\epsilon_b} + \frac{\gamma \langle z \rangle}{\epsilon_s},
\end{equation}
where $\langle z \rangle$ is measured from the upper interface, $\epsilon_s$ is the dielectric constant of the QW, and $\gamma$ is a numerical prefactor typically $0.5$--$0.7$.~\cite{Hampton1995}
In the following analysis, we used a linear fit function, like that in Fig.~3(a), to \textcolor{black}{determine $c_G$ as a function of $V_\text{FG}$.
This allows us to extract $c_Q$ from $c_\text{FG}$ using eq.~(\ref{c_FG}) without the need for knowledge of $\gamma$ and $\langle z \rangle$.
The linear fit agrees with $C_\text{exp}$ to within \textcolor{black}{$0.3$}~pF, with the corresponding errors in $c_Q$ and $\mu$ estimated to be less than $\textcolor{black}{\pm} 3\%$.
}

Figure~3(b) plots $c_Q$ obtained from the $C_\text{exp}$ data at $B=0$~T as a function of $V_\text{FG}$.
The $c_Q$ shows a sharp onset at $V_\text{FG}=\textcolor{black}{-0.8}$~V, where the 2DEG appears, and increases slowly with $V_\text{FG}$ at a nearly constant slope, except in the vicinity of the onset.
To check the validity of our analysis, we calculated $m^\ast$ from $c_Q$ using the relation $d n_s / d \mu = m^\ast/\pi \hbar^2$ and compared it with the existing theoretical models~\cite{Lin-Chung1993} for $m^\ast$ [inset of Fig.~3(b)].
The horizontal axis is the energy measured from the subband edge \textcolor{black}{$E_{1}$}.
The experimental data are plotted vs $\mu$ obtained using eq.~(\ref{mu}), with $\mu = 0$ defined as the middle point of the step edge in the $c_Q$ vs $\mu$ [see Fig.~5(a)].
The experimentally obtained $m^\ast$ increases with $\mu$, with the slope in \textcolor{black}{good} agreement with the model accounting for the nonparabolicity of the InAs conduction band (dotted line).
The quantitative agreement with theory is significantly improved when the effects of the $0.52$\% in-plane tensile strain in the InAs layer are taken into account (dashed line).~\cite{Yang1993,Lin-Chung1993}
The deviation from the linear dependence near the subband edge suggests electron-electron interaction that becomes important at low densities, which will be discussed later.
We emphasize, however, that the determination of $|g^\ast|$ using eq.~(\ref{mu}) is not affected by the value of $m^\ast$ obtained in our method, for it holds irrespective of the $m^\ast$ value.

\begin{figure}[ptb]
\includegraphics[width = 0.36\textwidth]{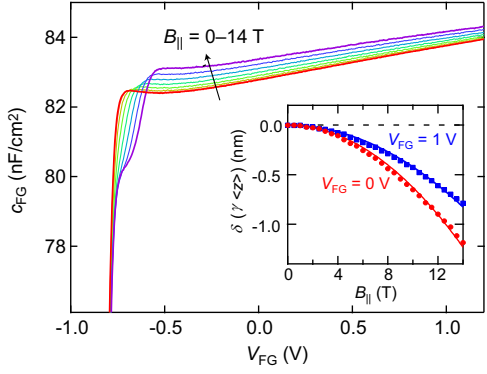}\caption{
$c_\text{FG}$ as a function of $V_\text{FG}$ measured at zero and various in-plane magnetic fields ($B_\parallel = 0$--$14$~T \textcolor{black}{with intervals of $2$~T}).
Inset shows the shift of the wave-function centroid with $B_\parallel$ estimated from the $B_\parallel$ dependence of $c_\text{FG}$ at $V_\text{FG} = 0$ and $1$~V (symbols).
Lines are quadratic fit.
}
\label{Fig4}
\end{figure}

Now we present results obtained with magnetic field $B_\parallel$ applied parallel to the 2DEG.
Figure~4 plots $c_\text{FG}$ vs $V_\text{FG}$ for $B_\parallel = 0$--$14$~T.
The data reveal that at large $B_\parallel$ an extra step feature develops near the onset, indicating the Zeeman splitting of the subband edge, as illustrated in Fig.~1(b).
As expected, the step becomes wider as $B_\parallel$ increases.
It is worth noting that, aside from the double-step feature, $c_\text{FG}$ is seen to vary with $B_\parallel$ also in the high-$V_\text{FG}$ range away from the subband edges.
This behavior, not expected from a simple picture, has previously been identified and explained as due to the $B_\parallel$-induced shift of the wave-function centroid in an asymmetric confinement potential.~\cite{Stern1968,Hampton1995,Jungwirth1995}
To quantify the shift, we translated the change in $c_\text{FG}$ at each $V_\text{FG}$ with respect to its $B_\parallel = 0$ value into $\gamma \langle z \rangle$, which we plot in the inset as a function of $B_\parallel$ for $V_\text{FG} = 0$ and $1$~V.
The shift can be fitted well with a quadratic function as $\delta(\gamma \langle z \rangle) = \alpha B_\parallel^2$.
The prefactor $\alpha$ is found to be $V_\text{FG}$ dependent, and can be expressed as $\alpha = (\textcolor{black}{6.20 - 2.05}V_\text{FG})\times 10^{-3}$~nm/T$^2$.
Once the centroid shift is absorbed in the $B_\parallel$ dependence of $c_G$, $c_Q$ in the high-$\mu$ region is no longer $B_\parallel$ dependent, as shown in Fig.~5(a), \textcolor{black}{where we plot $c_Q$ at $B_\parallel = 0$ and $14$~T as a function of $\mu$}.
\footnote{
\textcolor{black}{
The error in $\mu$ associated with the parabolic approximation of $c_\text{FG}$ vs $B_\parallel$ (Fig.~\ref{Fig4} inset) is negligible compared to that associated with the linear approximation of $c_G$ vs $V_\text{FG}$ [Fig.~\ref{Fig3}(a)].}
}
The double steps of $c_Q$ at $B_\parallel =14$~T are nearly equal in height, consistent with the equal DOS for the up and down spin states.
If we assume that the Fermi level aligns with the subband bottom at the middle of the step edges [crosses in Fig.~5(a)], we can deduce the energy difference $E_Z$ from their separation.
\textcolor{black}{
Note that the step edges are broadened by disorder, most likely due to background charged impurities, which limits the minimum resolvable $E_Z$ to $\sim 7$~meV in the present experiment.
}

\begin{figure}[ptb]
\includegraphics[width = 0.36\textwidth]{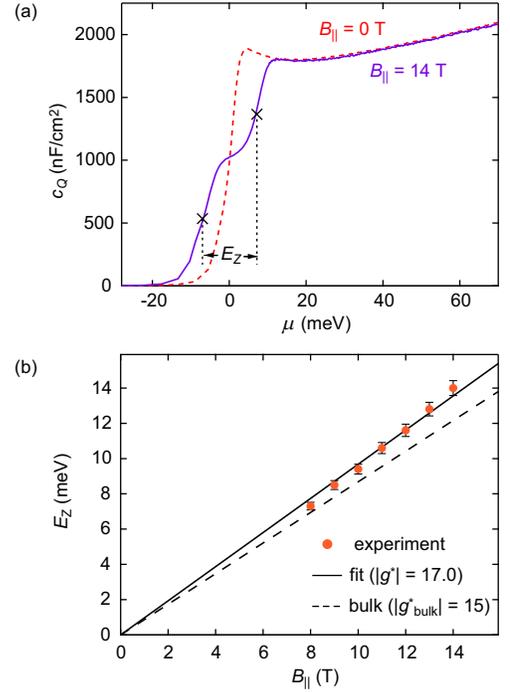}\caption{
(a) $c_Q$ at $B\textcolor{black}{_\parallel} = 0$ and $14$~T as a function of $\mu$.
Crosses indicate the middle of the step edges where the Fermi level is aligned to the subband edge.
$E_Z$ is deduced from the difference in $\mu$'s at these two points.
\textcolor{black}{
$\mu=0$ for the $B_\parallel = 14$~T data is set at the midpoint of these two points.
}
(b) $E_Z$ determined for various in-plane fields, plotted vs $B_\parallel$.
Open circles show experimental data.
\textcolor{black}{
The error bars indicate the error associated with the uncertainty ($\pm 3\%$) in $\mu$.
}
The solid line is a linear fit to the experimental data.
For comparison, $E_Z$ calculated using the $g$-factor of bulk InAs is shown by dashed line.
}
\label{Fig5}
\end{figure}

Figure~5(b) plots $E_Z$ determined for each $B_\parallel$.
Linear fitting of $E_Z$ vs $B_\parallel$ yields $|g^\ast|=\textcolor{black}{17.0 \pm 0.5}$.
Notably, this $|g^\ast|$ value is larger than that of bulk InAs ($g^\ast_\text{bulk}=-15$).~\cite{Pidgeon1967,J.Konopka1967}
In QWs, $g^\ast$ depends on the well width, reflecting the quantum confinement, which can be understood in terms of the energy dependence of $g^\ast$.
With the Kane model,~\cite{Kiselev1998}
\begin{equation}
g^\ast(E)=g_e - \frac{2E_P}{3}\frac{\Delta}{(E_g+E)(E_g+E+\Delta)},
\label{g-factor}
\end{equation}
where $E$ is the energy measured from the bottom of the bulk conduction band \textcolor{black}{$E_c$}, $E_g$ is the band gap, $\Delta$ is the spin-orbit splitting of the valence band, and $E_P = 2m_0 P^2/\hbar^2$ with $m_0$ the electron mass in vacuum and $P$ the Kane momentum-matrix element.
With the parameters for InAs ($E_g = 0.417$~eV, $\Delta = 0.39$~eV, and $E_P = 21.5$~eV),~\cite{Vurgaftman2001} eq.~(\ref{g-factor}) predicts that $g^\ast$ varies from $-14.6$ at the conduction-band bottom ($E = 0$) to, e.g., $g^\ast = -12.9$ at $E = 0.03$~eV.
The reported value $|g^\ast| \approx 13$ for a $15$-nm-wide InAs/AlSb QW, obtained using the coincidence technique,~\cite{Brosig2000} can therefore be explained by eq.~(\ref{g-factor}) if the Fermi level lies $\sim 0.03$~eV above the conduction-band bottom of bulk InAs.
However, quantum confinement cannot account for $|g^\ast|$ greater than the bulk value we observed.
We estimate $E \sim \textcolor{black}{0.03}$~eV in our $20$-nm-wide QW due to quantum confinement (\textcolor{black}{$\langle E_{1} - E_c(z) \rangle \sim 0.03$~eV}), which would yield $g^\ast = \textcolor{black}{-12.9}$.
As we argue below, the reduction of $E_g$ due to biaxial tensile strain can override the effects of $E$ on $g^\ast$.
If we take the energy of the light-hole band, which is at the top of the valence band for biaxial tension,~\cite{Lin-Chung1993} we find that the $0.52$\% strain decreases $E_g$ of InAs by $0.068$~eV (parameters are from Ref.~\onlinecite{Vurgaftman2001}).
Applying this value to eq.~(\ref{g-factor}) together with $E \sim \textcolor{black}{0.03}$~eV, we have $g^\ast = \textcolor{black}{-17.2}$.
Although a more elaborate theory, as that for $m^\ast$ in Ref.~\onlinecite{Lin-Chung1993}, is required for a rigorous discussion, our result suggests that the strain effect on $g^\ast$ is important in heterostructures as well as in nanowires and quantum dots;
it can be comparable to or even override the effects of quantum confinement for the case of tensile strain and therefore must be taken into account to discuss subtle effects, such as electron-electron interaction that becomes important at low density.~\cite{Tsukazaki2008}

We point out several differences between our method and the coincidence technique.
Firstly, the latter provides only the product $|g^\ast| m^\ast$; hence, an accurate determination of $|g^\ast|$ requires precise knowledge of $m^\ast$.
This is not the case for our method; as we demonstrated in this study, it allows for simultaneous and independent determination of $m^\ast$ and $|g^\ast|$.
Yet, our method requires large $|g^\ast|$ for the double-step feature to be resolved in the capacitance-voltage profile.
Secondly, in our method, $|g^\ast|$ is determied at a rather low electron density at which the upper-spin subband is depopulated and the system becomes fully spin polarized.
The corresponding $n_s$ depends on $|g^\ast|m^\ast$ and $B_\parallel$.
In the present case of InAs, where $|g^\ast|m^\ast/m_0 \sim 0.46$ leads to a rather large $n_s$ of $\sim 1 \times 10^{15}$~m$^{-2}$ at $B_\parallel = 14$~T, this value is still much smaller than the typical densities in coincidence experiments.
Thirdly, while the coincidence technique assumes $g^\ast$ to be isotropic, our method selectively measures the in-plane $g$-factor.
These differences will allow our method to provide information complementary to coincidence experiments and make it useful in some situations, such as for investigating the effects of electron-electron interaction on $m^\ast$ and $|g^\ast|$ in the absence of perpendicular magnetic fields.

\textbf{{Acknowledgements}}
The authors thank Hiroaki Murofushi for processing the device.
This work was supported by JSPS KAKENHI Grant No. JP15H05854.

%

\end{document}